\newcommand{\beq}{\begin{equation}}
\newcommand{\eeq}{\end{equation}}
\newcommand{\beqa}{\begin{eqnarray}}
\newcommand{\eeqa}{\end{eqnarray}}

\renewcommand{\lambda}{\ell}

\documentstyle[aps,prl,multicol,fancyheadings,epsfig,epsf,amsbsy,amssymb,amstex]
{revtex}
\begin{document}
\title{Unified Description of the Resonance Peak and Incommensuration
\\ in High-$T_c$ Superconductors}
\author{C.D. Batista, G. Ortiz, and A.V. Balatsky}
\address{
Theoretical Division, Los Alamos National Laboratory, Los Alamos, NM 87545}
\date{Received \today }

\maketitle

\begin{abstract}
We present a unified description of the resonance peak and low-energy
incommensurate response observed in high-$T_c$ cuprate superconductors.
We argue that both features have a purely magnetic origin and they
represent universal features of an incommensurate spin state both below
and above the superconducting transition temperature. In this
description the resonance peak is the reflection of commensurate
antiferromagnetism. Our theoretical scenario gives an account of the
main features observed in various families of superconductors and
predicts those not yet observed, like a resonance peak in
La$_2$NiO$_{4+x}$.
 
\end{abstract}

\vspace*{-1.0cm}
\begin{multicols}{2}

\columnseprule 0pt

\narrowtext

Neutron scattering experiments in the cuprates reveal two interesting
and seemingly unrelated effects: {\bf (a)} the low-energy
incommensurate peaks at momenta ${\bf k}=\left
(\frac{\pi}{a},\frac{\pi}{a}(1\pm \delta) \right )$ and $\left
(\frac{\pi}{a}(1\pm \delta),\frac{\pi}{a} \right )$ \cite{tranquada1}
with incommensuration $\delta$ (in units of $\frac{\pi}{a}$) and, {\bf
(b)} the so-called resonance peak at energies close to 40 meV in the
spin susceptibility \cite{resonance1} at the antiferromagnetic (AF)
vector ${\bf Q}=(\frac{\pi}{a},\frac{\pi}{a})$. These experimental
observations are widely believed to be important for our understanding
of the nature of the magnetic correlations and ultimately for the 
understanding of the superconductivity in high-$T_c$ materials. 

On the one hand the intensity and the energy $E_r$ of the resonance
peak seems to scale with the superconducting coherence energy scale
\cite{resonance2}
\beqa
E_r \simeq 5 \ k_B T_c \ .
\label{Er}
\eeqa
This experimental observation was in fact used in Refs.
\cite{resonance3} to relate the formation of the superconducting
coherence to the opening of a new spin-scattering channel in the
superconducting state that is impossible in the normal state above the
superconducting critical temperature $T_c$. Several theoretical
scenarios attributed the origin of the resonance peak to
superconducting coherence effects \cite{ressup} or considered it as the
fingerprint of a collective mode in an $SO(5)$ symmetric field theory
\cite{zhang}. 

On the other hand the direct proportionality between  $\delta$ and
$T_c$ has been observed in recent neutron scattering data for LSCO and
YBCO compounds \cite{YB}
\beqa
k_B T_c = 2 \hbar v^* \delta
\label{Tcdelta}
\eeqa
with some characteristic and material dependent velocity $\hbar v^*
\sim$ 17-35 meV \AA, where $\delta$ is measured in units of
$\frac{\pi}{a}$.

One could argue that there is no immediate connection between the two
phenomena \cite{review}. In this case any relationship to
superconductivity is accidental and hence there is no unifying physics
to be learned from comparing these two sets of observations. 
Alternatively one can attempt to prove that these two phenomena are
intimately related. This is the point of view we will advocate in this
article: We argue for the common origin of {\bf both the low energy
incommensurate response and resonance peak} as a magnetic  scattering
in the disordered incommensurate spin state. In our interpretation the
resonance peak is the spectral weight at energy $E_r$ associated to the
lowest energy spinon excitation with ${\bf k}={\bf Q}$ in a system with
spin incommensuration. A key aspect of our analysis is the realization
that the experimental observations are different manifestations of a
unique physical phenomenon. 

It is apparent that charge doping induces a certain spin ordering in
these low-dimensional, doped, antiferromagnets \cite{Note1} as a result
of competing interactions. Indeed, in one spatial dimension this can be
exactly shown to be the case \cite{our0}. Here, we will only consider
the spin degrees of freedom and show that, regardless of the nature of
the spin correlations, universal features emerge as a result of {\it an
arbitrary} magnetic incommensuration. A relevant question is, however,
up to what degree the charge channel affects the spin response and
therefore our conclusions. Since one spatial dimension is the case that
can be unambiguously addressed we have performed calculations on the
1$d$ $t$-$J$ model to show that our main thesis remains unchanged in
the presence of charges.

From Eqs. (\ref{Er}) and (\ref{Tcdelta}) one can conclude that
\beqa
E_r = \alpha \delta
\label{Erdelta}
\eeqa
with $\alpha \simeq (8-10) \hbar v^*$. This equation could indeed imply
that the resonance peak energy and incommensuration are directly
related. 
 

We find that there are direct experimental predictions based upon our
analysis that can help to resolve whether the incommensuration and
resonance are direct consequences of an incommensurate spin state. These
are:

\noindent
$\bullet$
Goldstone modes, characteristic of an ordered AF state, are shifted away
from ${\bf Q}$ by $\pm \frac{\pi}{a} \delta$ with a local maximum in the
dispersion relation $\omega_{\bf k}$ at ${\bf k}={\bf Q}$.

\noindent
$\bullet$
$E_r \propto \delta$, for small $\delta$. 

\noindent
$\bullet$
Resonance peak is associated with the lowest energy quasiparticle
excitation at ${\bf k}={\bf Q}$. 

\noindent
$\bullet$
Absolute intensity of the resonance peak decreases when:
the spin gap $\Delta_s$ (fixed $\delta$), or $\delta$
(fixed $\Delta_s$), or the temperature $T$ increases.

\begin{figure}
\epsfxsize=3.in
\centerline{\epsfbox{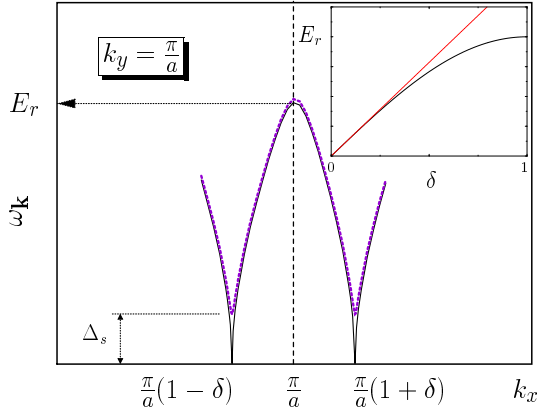}}
\caption[]{Spin-wave dispersion relation for an incommensuration in the
$x$-direction of $\pm \frac{\pi}{a} \delta$, i.e., ${\bf \tilde{Q}}=
\frac{\pi}{a}(1\pm\delta, 1)$. For completeness, we also display the
case where there is an spin gap $\Delta_s$ in the dispersion relation. 
For $\Delta_s$=0, we recover the massless Goldstone bosons at ${\bf
k}={\bf \tilde{Q}}$. The departure from linearity of the resonance
energy at ${\bf Q}$ is shown in the inset. }
\label{fig1}
\end{figure}

\begin{figure}
\epsfxsize=4in
\centerline{\epsfbox{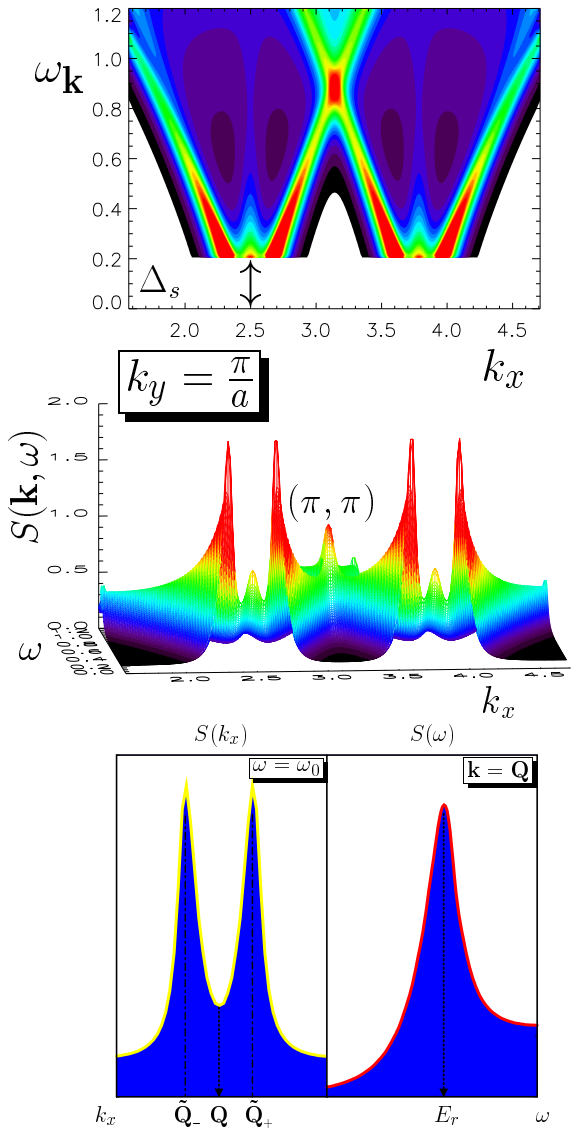}}
\vspace*{-0.0cm}
\caption[]{Energy and momentum dependence of the incommensurate spin
fluctuations as computed from case {\it (iii)}. The upper panel is a
contour plot of the lower one that clearly indicates the existence of a
resonance peak at ${\bf k} = {\bf Q}$. The intensity diminishes as one
moves away from the AF wave vector ${\bf Q}$. The modes at ${\bf k} =
{\bf \tilde{Q}}_{\pm}$ and $\omega \rightarrow 0$ are strongly affected
by the spin gap $\Delta_s$. As $\Delta_s$ increases there is a relative
increase of weight at the $(\pi,\pi)$-resonance. The lower panel
displays two cuts of the function $S({\bf k},\omega)$, one in momentum
and the other in energy ($\omega_0 < E_r$). The incommensurate vectors
are ${\bf \tilde{Q}}_{\pm} = {\bf Q} \pm \frac{\pi}{a} (\delta,0)$.} 
\label{fig2}
\end{figure}

We show that these features (illustrated in Figs.~\ref{fig1} and
\ref{fig2}) are universal, independently of the way the
incommensuration is established.

\vspace*{0cm}

To support such a claim we now present three different examples of
spin incommensuration (${\bf S}_i^2= S(S+1)$) with a common AF
background:

{\it (i)} The first example corresponds to ``striped'' topological
ordering, where on top of an AF backbone state $| AF \rangle$ there is
a ``ferromagnetic'' (FM) incommensurate exchange coupling $J'>0$ that
we model as
\begin{equation}
H' = -J' \sum_{\langle \alpha,\beta\rangle} {\bf S}_\alpha \cdot {\bf
S}_\beta \ ,
\end{equation}
where $\langle \alpha,\beta \rangle$ labels the FM link. The striped
arrangement of FM links induces $\pi$-shifted AF domains. A
straightforward perturbative argument leads to $E_r =
\frac{N_{L_{\;}}}{N} J' \langle AF | {\bf S}_\alpha \cdot {\bf S}_\beta
| AF \rangle \propto \delta J'$, where the ratio between the number of
FM links $N_L$ and lattice sites $N$ is $\delta$. 

{\it (ii)} The second is a spin-wave argument. Assume that on top of an
AF background there is a sinusoidally modulated spin structure such
that $\langle S^z_i \rangle = S \cos{({\bf \tilde{Q}} \cdot {\bf
r}_i)}$, with ${\bf \tilde{Q}}= {\bf Q} \pm (\frac{\pi}{a} \delta, 0)$.
This incommensurate spin structure can be stabilized by adding an
adequate term to a Heisenberg Hamiltonian \cite{liu}. When one
considers the simplest AF Hamiltonian 
the spin propagator $G_{ij}(t)=-i \langle \hat{T} S^+_i(t) S^-_j(0)\rangle$
has an equation of motion 
\begin{eqnarray}
[\omega^2+\omega_0^2] \ &G&({\bf k},{\bf k}';\omega) = \omega S \left [
\delta_{{\bf k},{\bf k}'+{\bf \tilde{Q}}} + \delta_{{\bf k},{\bf
k}'-{\bf \tilde{Q}}} \right]  \nonumber \\
&+& f({\bf k}-{\bf \tilde{Q}}) f({\bf k}-2{\bf \tilde{Q}}) \ G({\bf
k}-2{\bf \tilde{Q}},{\bf k}';\omega) \nonumber \\
&+& f({\bf k}+{\bf \tilde{Q}}) f({\bf k}+2{\bf \tilde{Q}}) \ G({\bf
k}+2{\bf \tilde{Q}},{\bf k}';\omega) 
\end{eqnarray}
with momentum-dependent frequency $\omega_0^2= -f({\bf k}) [f({\bf
k}+{\bf \tilde{Q}}) + f({\bf k}-{\bf \tilde{Q}})]$ and $f({\bf k})=
S[-J({\bf \tilde{Q}})+J({\bf k})]$, where $J({\bf k})$ is the Fourier
transform of the exchange interaction. This can be easily solved by using
either a continued fraction representation \cite{liu} of $G$ or mapping
its equation of motion to a Harper-like Hamiltonian \cite{ziman} that
can be diagonalized using standard methods. A simple analysis shows
that for small $\delta$, $E_r \propto \delta$ while Goldstone modes
appear at ${\bf \tilde{Q}}$. The spin-wave dispersion relation
$\omega_{\bf k}$ has a local maximum at ${\bf k}={\bf Q}$ (see
Fig.~\ref{fig1}).

{\it (iii)} Finally, one can use a Schwinger boson mean-field
description of an AF Hamiltonian, as in the spin-wave case, including a
term which gives rise to an incommensurate spin phase \cite{normand}.
Even though the Schwinger-boson approach does not explicitely break the
spin $SU(2)$ symmetry, as in the spin-wave approximation, both give the
same qualitative features for $\omega_{\bf k}$ and the dynamic magnetic
structure factor $S({\bf k},\omega)$.

From these examples one recognizes that certain features are
universal, regardless of the way the incommensuration is established.
In Fig.~\ref{fig1} we show the low-energy spin-wave dispersion relation
obtained following examples {\it (ii)} and {\it (iii)} which basically
lead to the same qualitative results. We observe the appearance of
Goldstone modes at the incommensurate wave vector ${\bf \tilde{Q}}$. 
The resonance energy $E_r$ (inset) is linear in $\delta$ for the values
of incommensuration observed experimentally. We have also considered
the case where there is a spin gap $\Delta_s$ in the excitation
spectrum \cite{Note2}. Note that the value of $E_r$ is not greatly
affected by the introduction of a spin gap. 

At low energies $S({\bf k},\omega)$ provides information on the
spectral weight of the lowest energy spin-quasiparticle excitations
(Fig~\ref{fig2}). In the presence of long range order, Goldstone bosons
appear at ${\bf k}={\bf \tilde{Q}}$ in $\omega=0$ due to the
spontaneously broken $SU(2)$ symmetry. The spectral weight, that is
proportional to the magnetic susceptibility $\chi({\bf k},\omega)$,
then diverges at ${\bf k}={\bf \tilde{Q}}$ and $\omega \rightarrow 0$. 
In the presence of a spin gap (as it seems to be the case for the
cuprates) these peaks become finite and indicate the presence of
short-range spin correlation at ${\bf k}={\bf \tilde{Q}}$. 

The low-energy peaks merge into a broad commensurate response at ${\bf
k} = {\bf Q}$ (Fig.~\ref{fig2}), where the resonance peak emerges as a
fingerprint of the incommensurate nature of the spin state. The
intensity of the resonance peak decreases with increasing $\delta$.
This can be understood as follows: The incommensurate phase is an AF
state modulated with a small $\delta$; therefore, in the presence of
incommensurability, there are still AF regions of size $\frac{2
a}{\delta}$ in the $x$ direction. Under these conditions we expect a
high intensity for the spectral weight of the lowest energy spin
excitation at ${\bf k}={\bf Q}$ which is reminiscent of the divergent
weight which is obtained when the ${\bf k}={\bf Q}$ channel is gapless
and the size of the AF region is infinite.  In the lower panel of
Fig.~\ref{fig2} we clearly show that both the low-energy incommensurate
response ($S({\bf k},\omega)$ for fixed energy $\omega_0$) and the
resonance peak ($S({\bf k},\omega)$ at ${\bf k} = {\bf Q}$) have the
same magnetic origin. 
\vspace*{-0cm}

\begin{figure}
\epsfxsize=4.0in
\centerline{\epsfbox{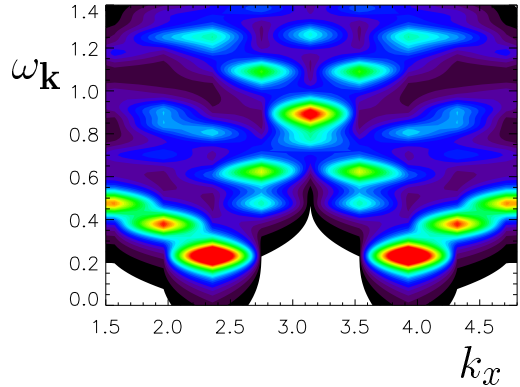}}
\vspace*{-0.0cm}
\caption[]{Same as upper panel in Fig.~\ref{fig2} but for a model with
both charge and spin degrees of freedom. The hole concentration is
$n_h=\frac{1}{4}$ and $J/t=0.4$. 
}
\label{fig3}
\end{figure}

We consider now the effect the charge degrees of freedom have on the
spin dynamics. To this end, we have calculated $S({\bf k},\omega)$ for 
a 1$d$ $t$-$J$ model (16 sites); the results are shown in
Fig.~\ref{fig3}. Clearly, the main universal features are still there:
Goldstone modes at the incommensurate points, $(\pi,\pi)$ peak at 
${\bf k} = {\bf Q}$, etc. The particle channel contributes to establish
the incommensuration (each charge carries an anti-phase boundary for
the AF order parameter \cite{our0}) but once it is established the main
qualitative features in the spin dynamics remain unchanged. This seems
to be the case for any doped AF Mott insulator whenever its state is
spin incommensurate. 
 \vspace*{-0.5cm}

These generic features have been observed in YBCO and Bi2212. In LSCO
the incommensuration has been certainly measured although there is no
indication of the presence of a resonance peak. However, this question
still has to be clarified experimentally, although it is possible that
a great degree of disorder in this compound is responsible for washing
out the peak and decrease its intensity. Our calculation indicates that
the relative weight of the resonance peak decreases and the weight at
the incommensurate points increases with decreasing spin gap
$\Delta_s$. This might explain why the resonance has not been observed
in LSCO where the spin gap, $\Delta_s \sim 7$ meV, is substantially
smaller than in YBCO. From our picture we predict a resonance peak at
$E_{r}=$14-16 meV for LSCO. Table \ref{table} summarizes the magnetic
properties of the different families of cuprate superconductors. Using
the linearity between $T_{c}$ and $\delta$, and the fact that $E_{r}$
is linear in $\delta$ for an incommensurate system (with small
incommensuration $\delta$), we can also explain the linear relation
between $E_{r}$ and $T_{c}$ measured in YBCO and Bi2212. Note that our
interpretation provides a definite meaning to the velocity $\hbar v^*$,
which is directly proportional to the rate of change of the resonance
energy $E_r$ as the result of changes in the spin incommensuration
$\delta$. 

Within our magnetic scenario one can qualitatively understand why in
the superconducting phase the intensity of the $(\pi,\pi)$ peak will be
enhanced, independently of the mechanism that drives the
superconducting state: Since holes moving in an uncorrelated fashion in
an incommensurate AF background carry an anti-phase domain wall,
whenever superconducting or pairing fluctuations coexist they will tend
to  ``annihilate'' those domains (holes form pairs) and make the system
more AF. These fluctuations produce some degree of disorder with
wavelengths longer than the incommensurate wavelength. The net effect
is a transfer of intensity in $S({\bf k},\omega)$ from the ${\bf
k}$-points near the incommensurate wavevectors ${\bf \tilde{Q}}_{\pm}$
towards the ${\bf k}$-points near the commensurate point ${\bf Q}$
(this is seen in calculations of the 1$d$ $t$-$J$ model). This scenario
is consistent with the temperature dependence of the incommensurate
magnetic response observed in La$_{1.86}$Sr$_{0.14}$CuO$_{4}$
\cite{review}.

To sharpen our argument about the purely magnetic origin of the
resonance peak in the incommensurate state we turn now to the {\em
non-superconducting} La$_2$NiO$_{4+x}$. This is a material where 
the existence of charged stripes is well established \cite{Tranquada2}.
From our model we expect that in the striped phase of LaNiO the 
magnetic incommensuration {\em coexists} with the resonance peak. We
estimate $E_r \sim 40-70$ meV for La$_2$NiO$_{4.13}$. It would be
interesting to see whether high-energy neutron scattering experiments
can be done on this material to search for a resonance peak in an
insulating material.
 
In conclusion, we argued that both the low energy incommensurate
response and resonance peak have a common magnetic origin. They are 
natural consequences of an incommensurate spin state. Superconductivity
is not the cause of the resonance peak although it could be the
consequence of incommensuration. There is an anomaly in the temperature
dependence of the resonance peak at $T=T_c$ with an abrupt change in
intensity above the level of a weaker, normal response. From our
description this fact can be related to the decreasing number of
anti-phase boundaries due to pairing fluctuations. These fluctuations
transfer spectral weight from the region around the incommensurate
${\bf k}$-points into the region near the commensurate ${\bf k} = {\bf
Q}$ point \cite{review}. In addition, the presence of a gap,
regardless of the superconducting mechanism, reduces the spin-particle
scattering increasing the antiferromagnetic fluctuations. $E_r$ emerges
as the result of the competition between antiferromagnetism (which can
lead to charge confinement) and delocalization (driven by kinetic
energy) therefore defining an additional characteristic energy scale.
From our findings, $E_r$ is intimately related to the emergent length
scale $\delta$. The purely magnetic origin of the resonance peak
explains why it has been observed above $T_{c}$. Most of the previous
descriptions of the origin of the $(\pi,\pi)$-resonance
\cite{ressup,resonance3,zhang} invoked coherent effects or collective
modes present in the superconducting state. These theories cannot
explain the many cases where the resonance has been observed above the
superconducting transition \cite{review}. An indirect confirmation of
our phenomenology is related to recent experimental observations on the
effect of nonmagnetic Zn impurities in YBCO \cite{sidis}. It is known
that Zn-doping suppresses superconductivity and at the same time leads
to a resonance peak above $T_c$, seemingly due to increased AF
correlations. 

Work at Los Alamos is sponsored by the US DOE under contract
W-7405-ENG-36. We also acknowledge useful discussions with P. Bourges,
P.C. Dai, I. Martin, H.A. Mook, D.K. Morr, J.M. Tranquada, and S.
Trugman.

\begin{table}[htb]
\begin{tabular}{l c c c c}
{\bf Compound} & LS$_{0.1}$CO & YBCO$_{6.85}$ & YBCO$_{6.6}$ & Bi2212  \\
\hline
\hline
$\hbar v^*$ (meV \AA) & 17 & 35 & 35 & 29  \\
\hline
$E_r$ (meV)& {\bf 14-16}& 41 & 34 & 37 \\
\hline
$\delta$ (r.l.u.) & 0.10 & 0.12 & 0.11 & 0.16 \\
\hline
$T_c$ ($K$) & 30 & 89 & 62.7 & 84 \\
\hline
Refs.& \cite{YB} & \cite{Bourges3,YB} & \cite{YB} &  \cite{BSCO,He} \\ 
\end{tabular}
\caption[table] {Properties of various families of high-$T_c$
compounds. Note that $E_r$=14-16 meV is our prediction for LS$_{0.1}$CO
based upon the linear relation between $E_r$ and $\delta$. Precise
values for $\hbar v^*$, $\delta$ and $E_r$ are taken from experimental
data. The proportionality $E_r \sim T_c \sim \delta$ holds within
10$\%$ accuracy.} 
\label{table}
\end{table}
 
\end{multicols}

\end{document}